\documentclass[11pt,draftclsnofoot,onecolumn]{IEEEtran}
\IEEEoverridecommandlockouts
\usepackage[utf8]{inputenc} 
\usepackage[T1]{fontenc}
\usepackage{url}              
\usepackage{cite}             
\usepackage[cmex10]{amsmath}
\usepackage{amssymb,amsmath,amsthm,latexsym,amscd,amsfonts,mathtools}
\usepackage{dsfont}
\usepackage{graphicx}         
\usepackage{textcomp}
\usepackage{xcolor}
\usepackage{subfigure}        
\usepackage{float}            
\usepackage[linesnumbered,ruled,vlined]{algorithm2e}
\usepackage[colorlinks=true, linkcolor=blue, citecolor=green]{hyperref} 
\usepackage{multirow}         
\usepackage{tikz}             
\usepackage{booktabs}         
\usepackage{stackrel}
\usepackage[top=0.75in,bottom=1in,left=0.625in,right=0.625in]{geometry}
\usepackage{tikz}
\usetikzlibrary{%
	arrows,%
	arrows.meta,
	shapes,
	shapes.misc,
	shapes.arrows,%
	backgrounds,
	calc,
	chains,%
	matrix,%
	positioning,
	scopes,%
	decorations.pathmorphing,
	shapes.geometric,
	shadows,
	decorations.pathreplacing,
	decorations.markings,
	patterns
}

\newtheorem{defi}{Definition}
\newtheorem{thm}{Theorem}

\newtheorem{corol}{Corollary}

\newcommand{\Ac}{\mathcal{A}}

\newcommand{\Cc}{\mathcal{C}}
\newcommand{\Dc}{\mathcal{D}}
\newcommand{\Ic}{\mathcal{I}}
\newcommand{\Lc}{\mathcal{L}}
\newcommand{\Uc}{\mathcal{U}}
\newcommand{\Pc}{\mathcal{P}}
\newcommand{\Xc}{\mathcal{X}}

\newcommand{\Mc}{\mathcal{M}}

\newcommand{\Nc}{\mathcal{N}}

\newcommand{\alphat}{\tilde{\alpha}}
\newcommand{\betat}{\tilde{\beta}}
\newcommand{\rhot}{\tilde{\rho}}
\newcommand{\etat}{\tilde{\eta}}
\newcommand{\gammat}{\tilde{\gamma}}
\newcommand{\gammabt}{\tilde{\breve{\gamma}}}
\newcommand{\Yt}{\tilde{Y}}

\newcommand{\nht}{\hat{n}}

\newcommand{\E}[1]{\mathbb{E} \left[ #1 \right]}
\newcommand{\EWR}[2]{\mathbb{E}_{#1} \left [ #2 \right ]}

\newcommand{\vari}[1]{\text{Var} \left[ #1 \right]}

\newcommand{\pr}[1]{\Pr \left\{ #1 \right\}}

\newcommand{\lrpar}[1]{\left( #1 \right)}
\newcommand{\lrbrk}[1]{\left[#1\right]}
\newcommand{\lrbrc}[1]{\left\{#1\right\}}
\newcommand{\lrvrt}[1]{\left\vert#1\right\vert}

\DeclarePairedDelimiter{\ceil}{\lceil}{\rceil}

\DeclareMathOperator*{\argmax}{\arg\!\max}
\DeclareMathOperator*{\qq}{q}

\DeclareMathOperator*{\betad}{\mathrm{Beta}}
\DeclareMathOperator*{\prdx}{par}

\begin{document}
	
	\title{Paradise of Forking Paths: Revisiting the Adaptive Data Analysis Problem}
	
	\author
	{
		Authors\\
		Sharif University of Technology, Tehran, Iran
		Email: \{Authors\}@ee.sharif.edu}
	
	\author{Amir Hossein Hadavi, Mohammad M. Mojahedian and~Mohammad Reza Aref\\
		
		Information Systems and Security Lab (ISSL)\\
		
		Department of Electrical Engineering, Sharif University of Technology, Tehran, Iran\\
		
		Email: \{ah{\_}hadavi\}@ee.sharif.edu,  \{m.mojahedian\}@gmail.com,  \{aref\}@sharif.edu
		
		\thanks{}}
	
	\maketitle

\begin{abstract}
		The Adaptive Data Analysis (ADA) problem, where an analyst interacts with a dataset through statistical queries, is often studied under the assumption of adversarial analyst behavior. To decrease this gap, we propose a revised model of ADA that accounts for more constructive interactions between the analysts and the data, where the goal is to enhance inference accuracy. Specifically, we focus on distribution estimation as a central objective guiding analyst's queries. The problem is addressed within a non-parametric Bayesian framework, capturing the flexibility and dynamic evolution of analyst's beliefs. Our hierarchical approach leverages Pólya trees (PTs) as priors over the distribution space, facilitating the adaptive selection of counting queries to efficiently reduce the estimation error without increasing the number of queries. Furthermore, with its interpretability and conjugacy, the proposed framework allows for intuitive conversion of subjective beliefs into objective priors and their effortless updates to posteriors. Using theoretical derivations, we formalize the PT-based solution as a computational algorithm. Simulations further demonstrate its effectiveness in distribution estimation tasks compared to the non-adaptive approach. By aligning with real-world applications, this structured ADA framework fosters opportunities for collaborative research in related areas, such as human-in-the-loop systems and cognitive studies of belief updating.
\end{abstract}

\IEEEpeerreviewmaketitle

\section{Introduction}
Exploratory Data Analysis (EDA) emphasizes understanding data through adaptive analyses before imposing formal assumptions about its underlying distribution \cite{tukey1962future, donoho201550}. By facilitating model-free statistics and interactive visualizations, it empowers researchers to uncover data features and structures. Despite its advantages, such as decreased bias compared to model-based approaches, recent studies have highlighted potential pitfalls. A key challenge is the ``garden of forking paths'' problem, as discussed in \cite{gelman2013garden,gelman2014statistical}. This issue arises from the adaptive nature of EDA, where analysts advance their exploration based on observations from previous analysis rounds. Such adaptivity resembles a branching process, with the potential for exponential growth in the number of analysis paths. Furthermore, EDA's flexibility may encourage researchers to persist in their explorations until notable patterns emerge within the dataset.  However, the generalizability of such discoveries may be questionable due to the rapid growth of exploration paths. In essence, EDA involves a trade-off between bias and variance. While it is recommended for addressing bias, it is also cautioned against due to the potential risk of increasing variance. This study addresses the question: what is the effect of adaptivity on the accuracy of distribution estimation compared to the non-adaptive case? Is it constructive, as Tukey proposed \cite{tukey1962future, donoho201550}, or destructive, as warned in \cite{gelman2013garden, gelman2014statistical}?

Recent theoretical research has investigated the accuracy guarantees of adaptive processes. Among these, a mainstream is the ``adaptive data analysis'' problem, first discussed by \cite{hardt2014preventing, dwork2015preserving} in sense of computational hardness results and feasibility guarantees, respectively. In ADA, the analyst $\Ac$ adaptively asks a sequence of queries from the answering mechanism $\Mc$, which owns a dataset $ S $ consisting of $n$ i.i.d. samples drawn from a distribution $P$ defined over a domain $ \Xc $. At the $i$-th round of the analysis, the query $ q_i $ is designed based on the history $(q_1, a_1, \dots, q_{i-1}, a_{i-1})$, where $ a_j $ is $ \Mc $'s empirical answer to $ q_j $ using $ S $. The queries are generally assumed to be selected from the family of statistical queries, defined in Section \ref{S_Pre_Def}.

In most works, since the action model of $ \Ac $ is unknown, a conservative worst-case scenario is assumed. Specifically, $ \Ac $ is modeled as having unbounded capability to pose adversarial queries designed to disrupt the generalizability of the results. Conversely, $ \Mc $ seeks an efficient answering mechanism that maximizes the number of ``query-answer'' rounds while maintaining the distributional accuracy of the answers. The accuracy is evaluated by $ (\epsilon,\delta) $-PAC criterion:
\begin{equation}\label{popacc}
	\pr{\max_i{ \lrvrt{\text{q}_i(P) - a_i } } \geq \epsilon} \leq \delta ,
\end{equation}
where the probability is taken over all sources of randomness in the model. In the non-adaptive scenario, where queries are selected upfront, Hoeffding's inequality and the union bound guarantee that $ \mathcal{O}(e^n) $ queries can be accurately answered using empirical means. However, this answering approach may become unstable in the adaptive setting \cite{dwork2015preserving}, due to the risk of rapid leakage of dataset information. In order to control the leakage rate, or equivalently stabilizing the algorithm, various mechanisms have been proposed in the literature, among which randomizing the answers by adding noise is the most popular.

Using such obfuscation mechanisms and a background of developed theories in algorithmic privacy \cite{dwork2014algorithmic}, numerous lower bounds  have been derived in ADA studies for the number of manageable queries \cite{dwork2015generalization, wang2016minimax, rogers2016max, bassily2016typicality, feldman2017generalization, feldman2018calibrating, esposito2019new, fish2020sampling, dagan2022bounded, blanc2023subsampling}. These bounds are typically of the order $ \mathcal{O}(n^2) $, with improvements largely focused on refining constants. One of the most recent and notable results is presented in \cite{dagan2022bounded}. By designing bounded-support noise distributions, this work narrows the gap between the best-known lower bounds and the tightest upper bounds established in earlier converse studies, including \cite{hardt2014preventing}, \cite{steinke2015interactive}, \cite{bun2017make}, and \cite{ullman2018limits}.

While these worst-case bounds provide theoretical safeguards for adaptive scenarios, their relevance to real-world applications remains debatable. For instance, an empirical study by \cite{roelofs2019meta} on a considerable number of ML competitions found that adaptivity has a limited impact on overfitting in these cases. Alternatively, several works have modified the primary ADA model by considering more realistic facts. In \cite{zrnic2019natural}, the assumption of unbounded adaptivity in $\Ac$ is replaced with two cognitive biases: recency bias and anchoring bias. These biases result in reduced worst-case dataset leakage. Also, \cite{mania2019model} explores task-specific similarities between classifiers, which introduce correlations between submitted queries and naturally constrain the maximum leakage rate. Additionally, \cite{feldman2019advantages} demonstrates that having multiple classes in a prediction problem can mitigate the worst-case error of adaptive attacks, highlighting the role of problem structure in reducing vulnerability. Overall, these works revise the assumption of $ \Ac $'s unlimited performance in formulating statistical queries or employing adaptivity to any extent. Rather than restricting $ \Ac $'s capabilities, another approach is proposed in \cite{elder2016bayesian, elder2016challenges} by enhancing $ \Mc $'s knowledge of $ P $. Specifically, in addition to the dataset $ S $, a Dirichlet prior is assumed for $ \Mc$. While these revised models provide valuable insights into the practice of ADA, they still regard $ \Ac $'s actions as a potential threat. Furthermore, \cite{hadavi2024overtheairfederatedadaptivedata} incorporates physical considerations by examining how answers from $ \Mc $ are communicated to $ \Ac $ over an analog channel.  It also studies a scenario involving multiple responders answering queries in a federated manner to $ \Ac $. Consequently, like the primary model, they establish bounds on the maximum number of tolerable queries.

Adopting the perspective of an analyst who seeks to enhance the generalizability of results, rather than intentionally or potentially disrupting them, fosters a more constructive approach to ADA. This perspective aligns with the approach advocated in our paper. Among the existing ADA literature, we identified only one work, \cite{rogers2020guaranteed}, that models $ \Ac $ in a constructive manner. In this model, the analyst not only submits queries but also provides his guesses for the answers along with corresponding confidence interval lengths.As noted in \cite{pu2018garden}, two general purposes can be considered in adaptive analyses: description and inference. The purpose considered in \cite{rogers2020guaranteed}, and generally in all the discussed ADA works, can be categorized as description. In contrast, we focus on the other main purpose, i.e., inference. It is noteworthy that focusing the interest on maximizing the duration of ADA is somehow counterintuitive, at least in some typical use cases. Normally, ADA can help the analyst progress iteratively based on prior explorations, which can be expected to yield fewer errors over the same duration of analysis compared to the non-adaptive case. 
This consideration is particularly relevant for inference tasks. Nevertheless, even for description purposes, providing more informative insights with fewer queries is often desirable. Moreover, inferencing a credible estimate of the distribution allows one to answer additional queries without reusing the dataset. Therefore, in our study of the problem, we take the opposite approach compared to the primary ADA framework—focusing on minimizing the number of queries.

Another questionable aspect of the primary ADA model is that $\Mc$ is required to take on the responsibility of guaranteeing population accuracy. When $P$ is unknown, especially after discarding the pessimistic view, expecting $\Mc$ to be the sole responsible party for generalizability seems unfair. In our proposed setup, $ \Mc $ has access to samples from $P$, while $ \Ac $ possesses prior knowledge about $P$. Under this framework,, $\Mc$ is obligated to answer accurately with respect to $S$, while $\Ac$ should optimally utilize $\Mc$'s responses in combination with his prior knowledge. We consider distribution estimation as the ultimate objective of $ \Ac $. We assume that $\Ac$ progresses in a Bayesian manner, based on a non-parametric structure. In our approach, we assume that $ \Ac $ progresses in a Bayesian manner within a non-parametric structure. Bayesianity formalizes the evolution of beliefs, while non-parametricity represents their flexibility within an overarching framework. The assumption of Bayesian progress is in agreement with many neuroscientific or cognitive studies, c.f. \cite{shams2022bayesian, sohn2021neural, austerweil2010learning, ullman2020bayesian}.

In summary, our contributions are as follows:
\begin{enumerate}
	\item Proposing a revised Bayesian model for ADA which is more realistic.
	\item Designing a hierarchical solution for the proposed model with an applicable interpretability.
	\item Elaborating the proposed approach as a versatile framework for statistical and cognitive studies.
	\item Demonstrating the model's efficiency as an adaptive distribution estimation scheme compared to the non-adaptive scheme.
\end{enumerate}

The paper proceeds with the required preliminary concepts in Section \ref{S_Pre_Def}. The model setup is introduced in Section \ref{MSET}, followed by an elaboration of the solution in Section \ref{ABDE}. Simulation studies are presented in Section \ref{S_Simul}, and the paper concludes in Section \ref{S_Concul}.

\section{Preliminaries and Definitions} \label{S_Pre_Def}
Throughout the paper, we refer to the analyst $ \Ac $ by male pronouns and to the answerer (data owner) $ \Mc $ by feminine ones.

\begin{defi}[Statistical queries] \label{StatQ}
	A statistical query (SQ) is a bounded function like $\qq : \Xc \rightarrow [a,b]$ which acts on single data samples and it's oracle answer with respect to $P$ is:
	\begin{equation}
		\qq(P) \triangleq \EWR{x\sim P}{\qq(x)}.
	\end{equation}
\end{defi}
An empirical estimation of $ \qq(P) $ can be attained from $ S $. A common, but not necessarily optimum estimation, is the empirical average $ \qq(S) \triangleq \frac{1}{n}\sum_{x \in S}{\qq(x)} $.

\begin{defi}[Counting queries]
	When in the definition \ref{StatQ}, $\qq$ is an indicator function on a subset of $\Xc$, we call it a counting query.  
\end{defi}

\subsection{Bayesian non-parametric models}
Consider a family of probability distributions\footnote{
	To avoid measure theoretic discussions, we simply limit our attention to the distributions which posses a valid density function almost everywhere. Equivalently, we work with the absolutely continuous distributions with respect to Lebesgue measure.
} 
$ \Pc$  on $ \Xc $ which can be characterized by $ \theta \in \Theta $. We refer to $ \Pc_\Theta $ as a probability model. 
When there is no finite-dimensional $ \Theta $ which can characterize $ \Pc$, it is known as a non-parametric family or model\footnote{When viewing $ \Pc $  as a collection of distributions, we call it a distribution family. When focusing on the specific way of characterizing the distributions, we refer to $ \Pc_\Theta $  (or simply $ \Pc $) as a probability model.	
} \cite{kleijn2022frequentist}. Such families have the potential to provide more flexibilities in statistical modelings, namely Bayesian modeling.  

\subsection{Non-parametric priors}
We mean by a ``non-parametric prior'', a hyper-distribution over a non-parametric probability model. Such hyper-distributions can be seen as stochastic processes with probability density functions (PDFs) as their realizations. For some references see \cite{hjort2010bayesian, kleijn2022frequentist, kleijn2012lectures}. Generally, we denote the hyper-distributions by $ \Pi $. In this paper, we work with Pólya trees, a hierarchical family of non-parametric priors with recently proved convergence rate results \cite{castillo2017polya}. 

\subsubsection*{Prior hyperparameter}
Similar to the discussion for probability models, a family of priors over $\Pc_\Theta$ can also be parameterized by hyperparameters $ \gamma \in \Gamma $. Hence, $ \Pi_{\gamma} $ and $ \Pi_\Gamma $ represent a specific prior and the prior family, respectively.

\subsection{Pólya trees} \label{S_PTs}
\begin{figure}[htbp]
	\centering
	\resizebox{0.9\columnwidth}{!}{ 
		\begin{tikzpicture}[
			level distance=2.5cm,
			sibling distance=4cm,
			every node/.style={fill=black!10, rectangle, rounded corners, draw=black, align=center},
			level 1/.style={sibling distance=10cm},
			level 2/.style={sibling distance=5cm},
			level 3/.style={sibling distance=2.5cm},
			edge from parent/.style={draw, -{Latex[length=3mm,width=2mm]}, thick},
			edge from parent path={(\tikzparentnode.south) -- ++(0,-5pt) -| (\tikzchildnode.north)},
			edge from parent/.append style={thick,draw=black},
			label style/.style={fill=black!0, rectangle, rounded corners, draw=none, inner sep=1pt}
			]
			\node {\large $I_{0}$ \\[-0.5ex] \large $M_{0} = 1$}
			child {node {\large $I_{10}$ \\[-0.5ex] \large $M_{10} = Y_{0}$}
				child {node {\large $I_{20}$ \\[-0.5ex] \large $M_{20} = M_{10} Y_{10}$}
					edge from parent node[left, label style, yshift=-0.5cm,xshift=-0.1cm] {\large $Y_{10} \sim \betad(\alpha_{10}, \beta_{10})$}}
				child {node {\large $I_{21}$ \\[-0.5ex] \large $M_{21} = M_{10}(1-Y_{10})$}
					edge from parent node[left, label style, yshift=-0.5cm,xshift=-0.1cm] {\large $1 - Y_{10}$}}
				edge from parent node[left, label style, yshift=-0.5cm,xshift=-0.1cm] {\large $Y_{0} \sim \betad(\alpha_{0}, \beta_{0})$}}
			child {node {\large $I_{11}$ \\[-0.5ex] \large $M_{11} = 1 - Y_{0}$}
				child {node {\large $I_{22}$ \\[-0.5ex] \large $M_{22} = M_{11}Y_{11}$}
					edge from parent node[left, label style, yshift=-0.5cm,xshift=-0.1cm] {\large $Y_{11} \sim \betad(\alpha_{11}, \beta_{11})$}}
				child {node {\large $I_{23}$ \\[-0.5ex] \large $M_{23} = M_{11}(1-Y_{11})$}
					edge from parent node[left, label style, yshift=-0.5cm,xshift=-0.1cm] {\large $1 - Y_{11}$}}
				edge from parent node[left, label style, yshift=-0.5cm,xshift=-0.1cm] {\large $1 - Y_{0}$}};
		\end{tikzpicture}	
	}
	\caption{Random probability mass allocation by Pólya Tree. The root and two first layers are shown.}
	\label{fig:PT}
\end{figure}
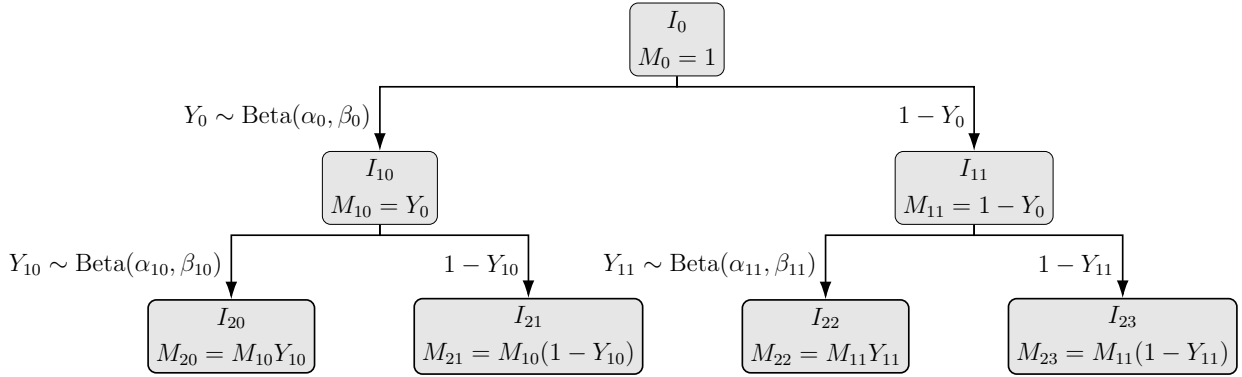

Pólya trees, as depicted in Fig.\ref{fig:PT}, define hierarchical, tree-structured, non-parametric priors over the space of PDFs. Consider an interval $ I_0 $ as the domain of the PDFs. Define $ \Ic_0 \triangleq \lrbrc{I_0} $. This set presents the root of the tree. In order to construct the first level of the hierarchy, divide $ I_0 $ in two subintervals $ I_{10} , I_{11} $. The division location can be arbitrary, e.g. at the midpoint. Define $ \Ic_1 \triangleq \lrbrc{ I_{10} , I_{11}} $. Continuing this procedure, the $ (l+1) $-th level of  hierarchy is derived from the $ l $-th level by dividing each element in $ \Ic_l $ in two subintervals. More exact, for each $ I_{ls} \in \Ic_l $ with $ s \in \lrbrc{0, 1, \dots, 2^l-1}  $, break $ I_{ls} $ in  two left and right subintervals and denote them by $ I_{(l+1)(2s)} $ and $ I_{(l+1)(2s+1)} $, respectively. Define the set $ \Ic_{l+1} $ as the collection of these new elements. It's clear that each level is a partition of $ \Xc $ which is finer than the previous level. We set $ \Ic \triangleq \bigcup_{l=0}^{\infty}{\Ic_l} $. This nested structure can be illustrated by a top-down binary tree. After the root, the parent of any node $ (l,s) $ is $ (l-1,\lfloor s/2 \rfloor) $ which we denote it by  $ \prdx(l,s) $. Random generation of density functions is done via random allocation of probability mass at any division. In detail, the probability mass of the interval $ I_{ls} $, denoted by the random variable $ M_{ls} $, is randomly shared between its left child $ I_{(l+1)(2s)} $ and its right child $ I_{(l+1)(2s+1)} $, according to the Beta distribution $ \text{Beta}(\alpha_{ls}, \beta_{ls}) $. More precisely,
\begin{align}
	&Y_{ls} \sim \betad(\alpha_{ls},\beta_{ls}), \label{Y_ls} \\
	&M_{(l+1)(2s)} = M_{ls}Y_{ls} \quad , \quad M_{(l+1)(2s+1)} =  M_{ls}(1-Y_{ls}), \label{cond_pls}
\end{align}
where $ Y_{ls} $'s are independent random variables. 
Realization of $ M_{ls} $ is done by starting at the root with $ p_0 = 1 $, and traversing the unique path to the node $ (l,s) $.  At each step, based on the movement direction, either $ Y_{l's'} $ or $ 1-Y_{l's'} $ is taken, where  $ (l',s') $ is the current node on the path.
Therefore, any specific realization of all the $Y$-variables, corresponds to a specific density function on $ I_0 $ (regardless of at most a countable number of points), and vice versa. Hence, $ \theta \triangleq y_{l=1 , s=0}^{\infty,2^l-1} $,  characterize the non-parametric family of density functions on $ I_0 $. 
Random generation of $ Y $-variables from distributions $ \betad(\alpha_{ls},\beta_{ls}) $,  yields a hyper-distribution on $ \Pc_\Theta $. 
Therefore, PTs constitute a prior family with the nested partitions $ \Ic $ and the hyperparameter(s) $ \gamma \triangleq (\alpha_{ls},\beta_{ls}) _{l=1,s=0}^{\infty,2^l-1}$. We refer to a specific PT by $\Pi_\gamma(\Ic) $ and the whole family by $\Pi_\Gamma(\Ic) $.

\section{Model setup} \label{MSET}

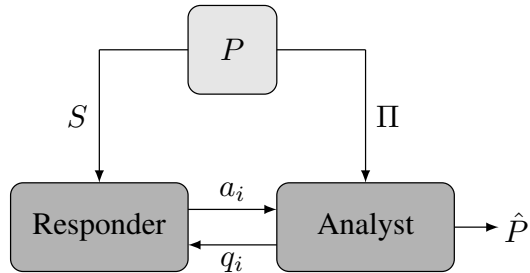
\begin{figure}[!b]
	\centering
	\resizebox {.4\columnwidth} {!} {
		\begin{tikzpicture}[scale=1]
			\tikzstyle{every node}=[font=\small]

			\draw[rounded corners,fill=black!10!white] (2,2.5) rectangle (3,3.5) node at (2.5,3) {$P$};
			
			\draw[rounded corners,fill=black!30!white] (0,0.5) rectangle (2,1.5) node [align=center,pos=.5] {Responder};
			
			\draw[rounded corners,fill=black!30!white] (3,0.5) rectangle (5,1.5) node [align=center,pos=.5] {Analyst};
			
			\draw[-latex] (2,1.2) -- (3,1.2) node at (2.5,1.4) {$a_i$};
			\draw[-latex] (3,0.8) -- (2,0.8) node at (2.5,0.6) {$q_i$};
			
			\draw (1,3) -- (2,3);
			\draw[-latex] (1,3) -- (1,1.5) node at (0.75,2.25) {$S$};
			
			\draw (3,3) -- (4,3);
			\draw[-latex] (4,3) -- (4,1.5) node at (4.25,2.25) {$\Pi$};

			\draw[-latex] (5,1) -- (5.5,1) node at (5.7,1) {$\hat{P}$};
			
		\end{tikzpicture}
	}
	\caption{Constructive interaction between the responder and the analyst aims to achieve a common goal (in this case, estimating the distribution $P$). The responder has access to a number of samples from the distribution $P$, while the analyst has prior knowledge about the distribution $P$.}
	\label{fig:ADA}
\end{figure}

In this section, we propose our formulation of the ADA problem and provide an overview of the solution scheme.  Detailed justification of the solution approach will appear in the Section \ref{ABDE}.
\subsection{Problem statement}
The proposed ADA model, featuring constructive responder-analyst interaction, is depicted in Fig. \ref{fig:ADA}. Same to the primary model, $\Ac$ interacts round-by-round with $\Mc$, who holds a dataset $ S $ of i.i.d. samples $ X_1^n $ drawn from $ P $, an unknown distribution for both $ \Ac $ and $ \Mc $. However, $ \Ac $, who can be considered as a domain expert, enjoys an overall belief about $ P $ which can be formulated as a prior $ \Pi $ on a probability model $ \Pc_\Theta $. He considers this prior for designing the queries, ‌combined with the answers received from $\Mc$. While such a prior is not assumed for $ \Mc $, her advantage is in having data samples from $ P $. The domain of $P$, denoted by $\Xc$, is assumed a bounded interval with the length $L$. Upon receiving an answer, $\Ac$ considers it as an empirical statistic of $ P $ and updates his prior on $ P $ to a posterior, by following the optimal Bayes' procedure. The ultimate objective of $\Ac$ is that for a fixed number of queries, attain the most credible estimation of $ P $. Equivalently, $ \Ac $ aims to minimize the number of queries to achieve a desired level of  accuracy in estimating $ P $. 

Conventional Bayesian distribution estimation assumes access to all data samples for a one-time update.  Our model explores a more complex scenario involving statistics of data (not raw samples) and an adaptive, sequential update process (not a one-time action). The first issue is a limitation and the second one is a facility for $ \Ac $. Here a fundamental question arises: How these two issues balances the ability of $ \Ac $ comparing to the conventional Bayesian case? 
Related studies for this scenario are scarce in the literature. However, the theoretical foundations of our solution scheme have already been established within non-parametric Bayesian theory\footnote{Note that in sequential methods like MCMC, the iterations primarily aim to address the challenge of computing the posterior, not necessarily for adaptivity purposes.}.

\subsection{Solution approach}
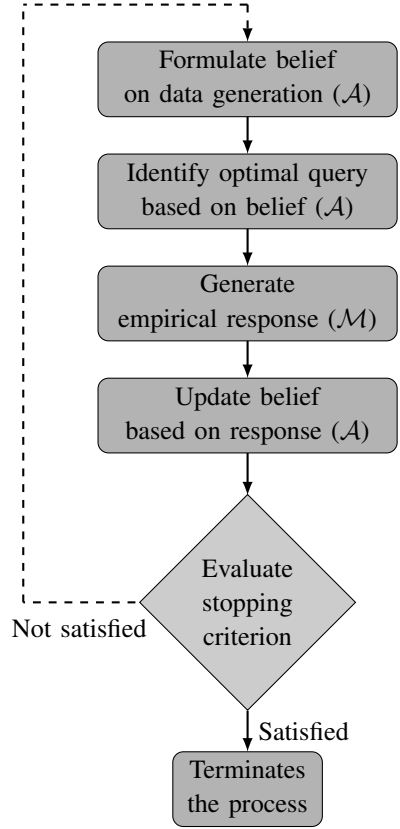
\begin{figure}[htbp]
	\centering
	\resizebox {.3\columnwidth} {!} {
		\begin{tikzpicture}[scale=1]
			\tikzstyle{every node}=[font=\small]

			\draw[rounded corners,fill=black!30!white] (-2,0) rectangle (2,1) node [align=center,pos=.5] {Formulate belief\\[-0.5em]on data generation ($\Ac$)};
			
			\draw[thick,-latex] (0,0) -- (0,-0.5);			
			
			\draw[rounded corners,fill=black!30!white] (-2,-1.5) rectangle (2,-0.5) node [align=center,pos=.5] {Identify optimal query\\[-0.5em] based on belief ($\Ac$)};
			
			\draw[thick,-latex] (0,-1.5) -- (0,-2);
			
			\draw[rounded corners,fill=black!30!white] (-2,-3) rectangle (2,-2) node [align=center,pos=.5] {Generate\\[-0.5em] empirical response ($\Mc$)};
			
			\draw[thick,-latex] (0,-3) -- (0,-3.5);
			
			\draw[rounded corners,fill=black!30!white] (-2,-4.5) rectangle (2,-3.5) node [align=center,pos=.5] {Update belief\\[-0.5em] based on response ($\Ac$)};
			
			\draw[thick,-latex] (0,-4.5) -- (0,-5.07);

			\node[diamond,draw,fill=black!20!white, minimum width = 2.5, minimum height = 3,align=center] at (0,-6.5) {Evaluate\\[-0.5em] stopping\\[-0.5em] criterion};
			
			\draw[thick,-latex] (0,-7.95) -- (0,-8.5);
			
			\draw[rounded corners,fill=black!30!white] (-1,-9.5) rectangle (1,-8.5) node [align=center,pos=.5] {Terminates\\[-0.5em] the process};
			
			\draw[thick,dashed] (-3,-6.5) -- (-3,1.5);
			\draw[thick,dashed] (-1.5,-6.5) -- (-3,-6.5);
			\draw[thick,dashed] (0,1.5) -- (-3,1.5);
			\draw[thick,dashed,-latex] (0,1.5) -- (0,1);
			
			\node at (0.75,-8.225) {Satisfied};
			\node at (-2.25,-6.85) {Not satisfied};

		\end{tikzpicture}
	}
	\caption{The overall flow of our proposed solution scheme can be illustrated as a five-step iterative process in each round.}
	\label{fig:ADA_flow}
\end{figure}

Given the prevalent role of human analysts in various daily applications, our focus is on scenarios where the analyst is human. Therefore, we aim for the solution to be compatible with such situations. A helpful observation from the problem statement is that the general procedure remains consistent in each round. Specifically, the overall process during each round is as follows:
\begin{enumerate}
	\item  Evoking the current belief on the data-generating rule (by $\Ac$)
	\item Asking the most useful query based on the belief (by $\Ac$)
	\item Providing an empirical response to the query, with a guaranteed level of fidelity to the dataset $ S $ (by $ \Mc $) 
	\item Optimally updating the belief considering the judgment about the answer (by $ \Ac $)
	\item If the stopping criterion is met, the process terminates; otherwise, the process reverts back to step 1 (by $\Ac$)
\end{enumerate}
The flowchart of this five-step iterative algorithm is shown in Fig. \ref{fig:ADA_flow}. To convert this procedure into an analytical algorithm, a mathematical interpretation of each step is required, which we address in the next section. Our methodology adopts a non-parametric approach to flexibly model the Bayesian evolution of the analyst's mind about the distribution. This can be viewed as a compromise between model-based and model-free approaches.  We choose PTs as non-parametric priors over the space of density functions due to their adaptablity in approximation with tunable precision. The intrinsic features of PTs align well with the inherent nature of the problem and our goal for the solution to be intuitively applicable to human analysts. Their hierarchical structure, accompanied by interpretable hyperparameters at each layer, facilitates the elicitation of the analyst's prior beliefs with tunable resolutions. Furthermore, their conjugacy property, coupled with the sufficiency of the counting queries for Bayesian updating, empowers $ \Ac $ to update the prior through indirect observations using straightforward computations.

\section{Adaptive Bayesian distribution estimation} \label{ABDE}
In this section, we delve into the details of the aforementioned five-step iterative procedure and convert it into an analytical algorithm.
\subsection{1st step: Evoking the current belief}
At the first step, a prior elicitation task is desired \cite{mikkola2021prior}, which in general can be challenging \cite{gelman2020bayesian}. It's not reasonable for a human analyst to run such task for each round of a real-time procedure. Therefore, it is better for the analyst to determine a general strategy for converting his subjective beliefs to analytic priors before starting the analysis. Fortunately, the top-down construction of PTs enables $\Ac$ to interpret his prior beliefs about $P$ in a hierarchical manner. In this structure, he can consider the division ratios through which the probability mass of parent nodes is distributed among their child nodes. He starts this process from the longest interval, i.e., $ I_0 = \Xc $, and then hierarchically proceeds to finer divisions, invoking his subtler beliefs. From \eqref{cond_pls}, we know $ Y_{ls} = \frac{M_{(l+1)(2s)}}{ M_{ls}}$. By defining $ \eta_{ls} \triangleq \alpha_{ls} + \beta_{ls} $ and $ \rho_{ls} \triangleq \frac{\alpha_{ls}}{\eta_{ls}} $, and considering \eqref{Y_ls},  we can write:
\begin{align}
	\E{Y_{ls} } &= \frac{\alpha_{ls}}{\alpha_{ls}+\beta_{ls}} = \rho_{ls}, \label{ExpY} \\
	\vari{Y_{ls}} &= \frac{\alpha_{ls} \beta_{ls}}{\lrpar{\alpha_{ls} + \beta_{ls}}^2 \lrpar{\alpha_{ls} + \beta_{ls} + 1}} = \frac{\rho_{ls}(1-\rho_{ls})}{1 + \eta_{ls}}. \label{VariY}
\end{align}
Hence, $ \rho_{ls} $, on average, represents $\Ac$'s belief about the division ratio at node $ (l,s) $. Moreover, for a constant $ \rho_{ls}$, a higher $ \eta_{ls} $ results in reduced variance of $Y_{ls}$, reflecting $\Ac$'s confidence level in his belief. 

\subsection{2nd step: Asking the most useful query}
This step involves additional details and will be explained through several subparts.
\subsubsection{On the usefulness criterion}
According to the problem statement, at each round, $\Ac$ should adaptively select a query based on the previous ``query-answer'' pairs and his prior knowledge of $P$. Since, after observing the answer, in the 4th step he updates the prior to the posterior in an optimal Bayesian way, all the extractable information in the answer is already embedded in the posterior. Therefore, in each new round, it is sufficient to consider only the current prior, which is equal to the last posterior from the previous round. Hence, in this step the main question is that based on the current prior, what is the most useful query?  Ideally, all the information of $ S $ about $ P $ is embedded in a (Bayesian) sufficient statistic. However, in non-parametric models there is no sufficient statistic which can be captured by a fixed number of statistical queries.  Thus, how should $\Ac$ select his adaptive queries to enhance their informativeness when they are considered as partial statistics? A clear answer, requires a utility criterion for informativeness. The choice of a utility function depends on the analysis goal, which in our problem aligns with the metric that evaluates the distribution estimation. In this study, we choose the mean squared error (MSE) metric. As a result, the ultimate objective is formulated as:
\begin{equation}\label{MSE_objective}
	\text{MSE} = \min_{\Pi}{\EWR{\Pi}{\int_{\Xc}{\lrvrt{P(x)-\hat{P}(x)}^2} dx}},
\end{equation}
where $ \Pi $ is the final resulted posterior from the analysis procedure. 

\subsubsection{Hierarchical scheme of adaptivity}
As it's also considered in the cognitive science literature \cite{drevet2022efficient, fuchs2023modeling, ullman2020bayesian}, a natural way to describe the adaptive procedure of learning is to use a hierarchical view. Using this view, we propose that $\Ac$ evolves hierarchically based on the described dyadic tree in subsection \ref{S_PTs}. In each round, he asks a counting query to know the sample counts of a sub-interval in $\Ic$, which corresponds to a node on tree. The choice of nodes is hierarchical; therefore, he chooses from nodes whose parent sample counts are already known. Without loss of generality, one can always query about the left child nodes to reveal the sample counts of both the left and right child nodes. See ``Step 4" below for more details. We denote the set of candidate nodes by $\Cc$.

\subsubsection{Analysis tree}
Due to the hierarchical progression of $\Ac$, at the end of each round, the nodes with unfolded sample counts constitute a finite subtree of the infinite dyadic tree.  We denote this subtree by $ T $. In this subtree the presence of nodes are necessarily pair-wise. In other words, for any node of $T$, his sister is also in $T$. In \cite{castillo2022optional}, trees with this property are named as ``full-binary trees''. However, in our work, we refer to them as ``analysis trees'' (AT) due to their natural emergence in our proposed ADA procedure. In general, $T$ can be asymmetric, which means that its leaves may be at various levels. We name the set of leaves of $T$ set by $ \Lc_T $. It is clear from the previous subpart that the set $\Cc$ is constituted by the left children of nodes in $ \Lc_T $. Each $ \Lc_T $ corresponds to a particular partition of $ \Xc $, which is denoted by $ \Ic_T $. 

\subsubsection{Finite Pólya trees }
Similar to PTs, based on any finite tree $T$, one can define a family of priors over a subspace of distributions with a resolution limited by the structure of the tree. We name such prior families as finite Pólya trees (FPTs), and denote them by $ \Pi_\gamma^T(\Ic) $, where $\Ic$ emphasizes the relation to the full nested structure of subintervals. Furthermore, $ \gamma $ represents the collection of hyperparameters $(\alpha_{ls},\beta_{ls})_{(l,s)\in T^-}$, where $T^- \triangleq T - \Lc_T$.  
\\
Asking one more counting query and receiving the answer allows $ \Ac $ to derive a refined partition from $ \Ic_T $. For this, after $ \Ac $ converts his beliefs about the nodes in $\Lc_T$ into Bayesian priors according to the first step, he needs to select the best candidate for querying, which is the focus of the next subpart. Once the query-answer process on the selected node is complete, the parent of the queried node branches into its child nodes, i.e., the queried node and its sister. Meanwhile, the prior $(\alpha, \beta)$ hyperparameters of the parent node are updated to posterior values based on the unfolded sample counts (see Step 4).

\subsubsection{Deriving the utility function for the node selection }
To determine which leaf should be selected for branching, we consider $ T^+ $, a tree derived from $ T $  by branching all its leaves into their respective child nodes. Then, using \eqref{cond_pls}, we analyze which pair of child nodes, when their probability mass generation is altered from the prior distribution to the posterior distribution, results in a greater reduction in the MSE.

Any partition $ \Ic_{T^+} $, induces a piece-wise constant approximation of $P$ by considering mass probabilities that $P$ assigns to any $I \in \Ic_{T^+}$. We denote this approximation by $P_{T^+}$ and the mass probability that $ P $ assigns to $I$ by $p_I$. Correspondingly, a stochastic estimation of $P_{T^+}$ can be drawn from $ \Pi_\gamma^{T^+}(\Ic) $, which we denote it by $\hat{P}_{T^+}$. Now, based on $T^{+}$ an approximation of \eqref{MSE_objective} can be written as
\begin{align}\label{approx_MSE}
	\hat{\text{MSE}} &= \EWR{\Pi_\gamma^{T^+}(\Ic)}{\int_{\Xc}{\lrvrt{P_{T^+}(x)-\hat{P}_{T^+}(x)}^2 dx}} \nonumber \\
	&= \EWR{\Pi_\gamma^{T^+}(\Ic)}{\sum_{I \in \Ic_{T^+}}{\lrpar{\frac{p_I}{L_I}-\frac{M_I}{L_I}}^{2} \times L_I}} \nonumber \\
	&= \sum_{I \in \Ic_{T^+}}{L_I^{-1} \times \E{{\lrpar{p_I-M_I}^2}} } \nonumber \\
	&= \sum_{ I \in \Ic_{T^+}}{\frac{2^{l_I}}{L} \times \lrbrc{\lrpar{p_I-\nu_{M_I}}^2 + \E{\lrpar{\nu_{M_I}-M_I}^2}}}, 
\end{align}
where $L_I$ and $l_I$, represents length and level number of $I$, respectively. Also, $\nu_{M_I}$ is the mean of $M_I$. In \eqref{approx_MSE}, the first term in summation represents the bias and the second term represents the variance of $\hat{P}_{T^+}(x)$ in estimating $P_{T^+}$. The crucial question now is how well $ \hat{\text{MSE}} $ represents $ \text{MSE} $. 
Generally, we can write: $ 	\text{MSE} = 	\hat{\text{MSE}} + \text{Residual}$.  The residual part depends on the complement of ${T^+}$ with respect to the full infinite tree. Since the progress in our proposed ADA analysis procedure is hierarchical, reducing the residual part could not be the direct concern of the current round. So, $\Ac$ can focus on the reduction of $ \hat{\text{MSE}} $.

As illustrated in \ref{S_PTs},  for any $I \in \Ic_{T^+}$, $M_I$ is generated by multiplication of i.i.d. Beta R.V.s belonging to the nodes on the path from the root to $\prdx(I)$. For all of these nodes, except $\prdx(I)$, $(\alpha,\beta)$ hyperparameters have been already updated to their posterior values. Now, in this round the specific question is that updating the hyperparameters of which leaf of $ T $ yields to more reduction in \eqref{approx_MSE}.  Choosing any node in $ \Lc_T $ like $ b $, to query its left child and consequently update $ (\alpha_b,\beta_b) $ , affects two terms of the series in \eqref{approx_MSE} which belong to $ b $'s children. We denote this part of series by $E(b)$. By indicating $b$'s left and right child nodes by subscripts $L$ and $R$, we can write:
\begin{align}\label{leaf_quota}
	\frac{L E(b)}{2^{l_b}}&= (p_L-\nu_b\rho_b)^2 + (\nu_b\rho_b-M_bY_b)^2 \nonumber \\ & \qquad +(p_R-\nu_b(1-\rho_b))^2 + (\nu_b(1-\rho_b)-M_b(1-Y_b))^2 \nonumber \\
	&\!\!\!\!\!\!\!\stackrel{(\nu_b\approx M_b)}{\approx}
	(p_L-\nu_b\rho_b)^2 + \nu_b^2\vari{Y} \nonumber
	\\ & \qquad + (p_R-\nu_b(1-\rho_b))^2 + \nu_b^2\vari{1-Y} \nonumber \\
	&\stackrel{\eqref{VariY}}{=}  (p_L-\nu_b\rho_b)^2 + (p_R-\nu_b(1-\rho_b))^2 + \frac{2\nu_b^2\rho_b(1-\rho_b)}{1+\eta_b} ,
\end{align}
where $\nu_b \approx M_b$ is due to the fact that distributions of $b$'s ancestors have been already updated to the posterior and consequently, $\vari{M_b} \approx 0$. Also, we denote \eqref{leaf_quota} by $\frac{\hat{E}(b)}{2^{l_b}}$.  Now, we can derive the desired utility function for any $I \in\Cc$ by considering the amount of reduction in $\hat{E}(\prdx(I))$ before and after posterior update. We define:
\begin{equation}\label{var_diff}
	v(I) = \hat{E}_\text{prior}(\prdx(I)) - \hat{E}_\text{posterior}(\prdx(I)).
\end{equation}
From \eqref{VariY} and \eqref{conjpost} we have:
\begin{align}
	\vari{\Yt_{\prdx(I)}} &= \frac{\rhot_{\prdx(I)}(1-\rhot_{\prdx(I)})}{1 + \etat_{\prdx(I)}}, \label{VRF}\\
	\etat_{\prdx(I)} &= \eta_{\prdx(I)} + n_{\prdx(I)}, \label{Eta_t}\\
	\rhot_{\prdx(I)} &= \frac{\rho_{\prdx(I)}\eta_{\prdx(I)} + n_I}{\etat_{\prdx(I)}}, \label{Rho_t}
\end{align}
where $\sim$ sign means belonging to the posterior distribution. Remember that the nodes in $\Cc$ are the left child nodes of the nodes in $\Lc_T$. Hence, in \eqref{Eta_t} $n_{\prdx(I)}$ is known and in \eqref{Rho_t} $n_I$ is unknown. By plugging-in the mean-value estimator $ \nht_{I} = \rho_{\prdx(I)}n_{\prdx(I)} $ in \eqref{Rho_t}, we have:
\begin{equation} \label{Rho_t_est}
	\hat{\rhot}_{\prdx(I)} = \frac{\rho_{\prdx(I)}(\eta_{\prdx(I)} + n_{\prdx(I)})}{\etat_{\prdx(I)}} = \rho_{\prdx(I)}.
\end{equation}
Now, by using \eqref{Eta_t} and \eqref{Rho_t_est} in \eqref{leaf_quota} for the posterior, we will have:
\begin{align}
	\frac{\hat{v}(I)}{2} &= 2^{(l_{\prdx(I)}+1)}\nu_{\prdx(I)}^2\rho_{\prdx(I)}(1-\rho_{\prdx(I)}) \times  \lrbrk{\frac{1}{1+\eta_{\prdx(I)}}-\frac{1}{1+\eta_{\prdx(I)}+ n_{\prdx(I)}}}. \label{v_hat}
\end{align}
The calculation of \eqref{v_hat} depends entirely on $\prdx(I)$, rather than on $ I $ itself. Thus, we can equivalently consider the utility criterion solely in terms of the parent node. According to this criterion, we select the parent of the node associated with the new query. Consequently, we can redefine the set $ \Cc $ to consist of the parents of the candidates for the next query, which allows us to set $ \Cc \equiv \Lc $. In Algorithm \ref{Alg1}, presented at the end of this section, we apply these considerations. Finally, after eliminating the constants, our proposed utility criterion for selecting a counting query with near-optimal informativeness is formulated as 
\begin{align}
	u(I) \triangleq \frac{ 2^{l_{I}} n_{I} \nu_{I}^2\rho_{I}(1-\rho_{I}) }{ (1+\eta_{I})(1+\eta_{I}+ n_{I}) }, \label{u_1}
\end{align}
where $ I \in \Lc_T $. 

\subsection{3rd step: Empirical response to the query }
Following our revision on ADA in the introduction, in our model $ \Mc $'s accountability is to answer with a guaranteed level of fidelity with respect to $ S $, her only source of information about $ P $. Different fidelity approaches can be specified. A conventional choice is $(\epsilon,\delta)$-empirically-PAC answering, which is defined similar to \eqref{popacc}, by replacing $\text{q}_i(P)$ with $\text{q}_i(S)$. Based on her desired privacy level, $ \Mc $ can promise for proper values of $ (\epsilon,\delta) $. In this work, we assume that privacy is not a concern for $ \Mc $, and we set $ \epsilon = \delta = 0 $. Hence, $ \Mc $ responds solely with the empirical mean $\text{q}_i(S) $ without any additional disturbances. A natural use case for this assumption is when $ \Ac $ interacts fully with a large dataset through statistical queries. This aligns with EDA framework which is described in the introduction section.

\subsection{4th step: Bayesian belief update} 
Is it manageable for $ \Ac $ to optimally update his subjective beliefs about $ P $ using intuitive rules? Equivalently, can he apply 
Bayes' formula  through light computations to change his prior on $ P $ to a posterior? The answer depends on the structure of the prior. Particularly, it is yes, if $\Ac$ deals with a conjugate prior.  In context of our problem, conjugacy needs to be established with respect to the statistics received as query responses, rather than with the entire dataset.

Below, we state in Theorem \ref{FPTconj} that FPTs $ \Pi_\Gamma^T(\Ic) $ satisfy these requirements, provided the queries are selected from the counting queries $ \qq_I $ where $ I \in \Ic $. This theorem shows that the hyperparameters of parent nodes are updated based on how their left and right children inherit their samples. While the conjugacy of (infinite) Pólya Trees is a well-known result \cite{ferguson1974prior}, here we explicitly state and prove the conjugacy for FPTs. This clarification better illustrates the inductive and hierarchical nature of our proposed framework \footnote{It is worth noting that after completing our proof of Theorem \ref{FPTconj}, we discovered that a result closely related to this theorem was also recently proven in \cite{castillo2022optional}. However, while our theorem provides results based solely on sample counts, the referenced theorem assumes that all samples are fully given. Additionally, our inductive proof aligns well with the hierarchical approach we propose for constructing PTs. Nonetheless, we would like to emphasize that our proof was developed independently and addresses the specific statement required for this paper.}. Theorem \ref{FPTconj} is followed by \ref{PartialConj}, where we demonstrate a partial conjugacy version for updating FPT priors when only the answer to the most received counting query is available,  and $ \Ac $'s progression in previous rounds is due to our proposed procedure. Specifically, we describe how the hyperparameters of a leaf node of $ T $ can be updated through a single counting query about its left child node.

\begin{thm}[Conjugacy of FPTs] \label{FPTconj}
	Let $ T $ be an analysis tree as introduced in the second step. Assume that the FPT prior $\Pi_\gamma^{T}(\Ic) $ with $ \gamma = (\alpha_{ls},\beta_{ls}) _{(l,s)\in T^-}$ is given. Let $n_{ls}$ be the sample count of $ I_{ls} $ for any node $(l,s) \in T$. Then, the posterior $ \Pi(\cdot \mid (n_{ls})_{(l,s)\in T}) $ will be again a FPT on $ T $ with hyperparameters $ \gammat = (\alphat_{ls},\betat_{ls}) _{(l,s) \in T^-} $ for which:
	\begin{equation} \label{conjpost}
		\alphat_{ls} = \alpha_{ls} + n_{(l+1)(2s)} 
		\quad , \quad
		\betat_{ls} = \beta_{ls} +  n_{(l+1)(2s+1)}.
	\end{equation}
\end{thm}

\begin{proof}
	Recall that $ \Pi_\gamma^{T}(\Ic) $ determines a Bayesian prior on the subspace of PDFs with resolution limited to $ \Ic_T $. Denote this subspace by $ \Dc^{T} $.  Consider any $ D \in \Dc^{T} $. In the following, we use this fact that $ D $ is completely characterized by $ (d_I)_{I\in \Ic_T} $, where $ d_I \triangleq \int_{I}{D(x)dx} $. We prove the theorem by induction on the size of $ \Lc_T $. Note that $ \vert \Lc_T \vert  \geq 2$ with equality only for the case of the root and its two children. So, as the base case for induction, let $ T $ be this tree. In this case, $ D $ is simply a Bernoulli distribution with parameter $ d \in \lrbrk{0,1} $ on which the prior $ \Pi_\gamma^{T}(\Ic) \equiv \betad(\alpha,\beta)  $ is given. Hence, we have: $\Pi_\gamma^{T}(D;\Ic) = \betad(d;\alpha,\beta) = \frac{1}{\mathrm{B}(\alpha,\beta)} d^{\alpha-1}(1-d)^{\beta-1}$. From Bayes' formula we have: 
	\begin{align}
		\Pi(D \mid n_{10},n_{11})  &=  \frac{ \pr{n_{10},n_{11} \mid D} \Pi_\gamma^{T}(D) }{\int_{D'\in\Dc^{T}}{\pr{n_{10},n_{11} \mid D'} \Pi_\gamma^{T}(D')}}  \nonumber \\
		&\stackrel{\text{(a)}}{=} c_1 d^{n_{10}} (1-d)^{n_{11}} d^{\alpha-1}(1-d)^{\beta-1}  \nonumber \\
		&\stackrel{\text{(b)}}{=} \betad(d;\alpha+n_{10},\beta+n_{11}).
	\end{align}
	In (a), the constant $c_1$ does not depend on $D$. Furthermore, since the left-hand side of (b) —which must represent a distribution—is proportional to the beta distribution on the right-hand side, they should be equal. This completes the proof for the base. Also we simplified the notation  $\Pi_\gamma^{T}(D;\Ic) $ as $\Pi_\gamma^{T}(D)$.
	\\
	Now, assume that the claim is true for any analysis tree with $m$ leaves and let $T$ be a tree with $\lrvrt{\Lc_T }= m+1$. Arbitrarily select one of the nodes in $\Lc_T$ with maximum level number and denote it by $L$. Since $T$ is an analysis tree, $L$'s sister is also in $T $. Denote it by $R$. Since $L$ is selected with maximum level number, $R$ is also in $\Lc_T$. Otherwise, $R$'s child will be in $T$ with a greater level number than $L$. We refer to their parent with ``$\prdx$''. Without loss of generality, assume that $L$ is the left child. Also, define $\breve{T} \triangleq T - \lrbrc{L,R}$. Correspondingly, consider $ \breve{D} $ as $ D $ when limited to $ \breve{T} $, $ \breve{\gamma} \triangleq (\alpha_{ls},\beta_{ls}) _{(l,s)\in T^{-} - \lrbrc{\prdx} }$ and $ \gammabt $ its posterior update as \eqref{conjpost}.
	
	\begin{align}
		\Pi\lrpar{ D \mid (n_{ls})_{(l,s) \in T} } &  \stackrel{(c)}{=}  c_2   \pr{ n_L,n_R \mid D ,  (n_{ls})_{ (l,s) \in \breve{T} } } \Pi\lrpar{ D \mid (n_{ls})_{ (l,s) \in \breve{T} } } \nonumber \\
		& \qquad  \stackrel{\text{(d)}}{=} c_2 \pr{ n_L,n_R \mid D, n_{\prdx} } \Pi\lrpar{ (d_I)_{I \in \Lc_{\breve{T}}} \mid (n_{ls})_{ (l,s) \in \breve{T} } } \times \Pi\lrpar{ d_L, d_R \mid d_{\prdx} }  \nonumber \\
		& \qquad \stackrel{\text{(e)}}{=} c_2 \frac{ d_{L}^{n_L} d_{R}^{n_R} }{d_{\prdx}^{n_L+n_R}} \times \Pi_{\gammabt}^{\breve{T}} (\breve{D}) \times c_3 {(\frac{d_L}{d_{\prdx}})}^{\alpha_{\prdx}} {(\frac{d_R}{d_{\prdx}})}^{\beta_{\prdx}} \nonumber \\
		& \qquad  = c_4  \betad{ (\frac{d_L}{d_{\prdx}} ; \alpha_{\prdx} + n_L , \beta_{\prdx} + n_R) } \times \Pi_{\gammabt}^{\breve{T}} (\breve{D}) \nonumber \\
		& \qquad \stackrel{(f)}{=} \Pi_{\gammat}^{T} (D).
	\end{align}
In (c), we used the fact that $ \pr{x \mid y, z} = \frac{ \pr{y \mid x, z} \pr{x \mid z }}{ \pr{y \mid z } } $, where $ x \equiv D $, $ y \equiv (n_L, n_R) $, and $ z \equiv (n_{ls})_{ (l,s) \in \breve{T} }$. In (d), we considered the top-down Markovian random generation of mass probabilities. In (e), we applied the induction assumption for $ \breve{T} $, since $ \lrvrt{\Lc_{\breve{T}}} = m $. Also, the first term is simply a conditional probability, and the third term is due to random division of  $ d_{\prdx} $ into $ d_L $ and $ d_R $ according to the distribution $ \betad(\alpha_{\prdx}, \beta_{\prdx}) $. In (g), we considered the hierarchical random generation of mass probabilities as illustrated in \ref{S_PTs}. Additionally, a similar argument to the one provided for (b) is applied here.
\end{proof}

\begin{corol} \label{PartialConj}
	Assume that $ \Ac $ has progressed hierarchically in the previous rounds and obtained an analysis tree $ T $. Also, assume that he already has updated $ \gamma $ to $ \gammat $ using \eqref{conjpost}. Let ``$ \prdx $'' be a node in $ \Lc_T $ and let $ L $ be its left child. Then, in the new round, if $ \Ac $ query on $ L $ and receive the answer $ n_L $, it suffices for the optimal Bayesian update to only change $ ( \alpha_{\prdx} , \beta_{\prdx} ) $ to $ ( \alpha_{\prdx} + n_{L} , \beta_{\prdx} + n_{R} ) $, where $ n_R = n_{\prdx} - n_L $.
\end{corol}
\begin{proof}
	It immediately follows from the inductive proof scheme of Theorem \ref{FPTconj} that tree hyperparameters can be updated step-by-step after each new answer is received. As noted before, a node $ (l,s) $ is queried only when the sample count of $ \prdx(l,s) $ is already known.  Therefore, the sample count of its sister node is also unfolded as $ n_{\prdx(l,s)} -n_{ls}$ and the situation for using \eqref{conjpost} is prepared. 
\end{proof}

\subsection{The fifth step: Termination decision} 
In this study, since we assume a fixed budget for the number of queries, the analysis stops as soon as $k$ queries are asked. In more general case, the stopping criterion can be set based on the estimation precision. A reasonable choice is considering the utility function we derived in 2nd step. By this choice, the procedure terminates once the expected amount of variance reduction lies below a predefined threshold. 

\subsection{Algorithm formulation}
Finally, Our proposed scheme is presented in Algorithm \ref{Alg1}.
\begin{algorithm} \label{Alg1}
	\caption{Adaptive data analysis}
	\KwData{For $ \Ac $: Size of dataset $ n $, Number of queries $ k $ , PT prior $ \Pi_\gamma(\Ic) $ \\ 
		\qquad\quad\!\!\!\! For $ \Mc $: Dataset $ S $}
	\KwResult{PT posterior $\Pi_{\gammat}(\Ic)$}
	\textbf{Initialization:} Set $ \Cc = \lrbrc{I_0} $, $ n_{I_0}=n $, $ i = 0 $ ;
	
	\While{$ i<k $}{
		Evoke the prior belief for the newly added candidates in terms of $ (\rho, \eta) $; \\
		Set $ I_{\prdx} = \argmax_{I\in\Cc}{u(I)} $, and $ I_L$ and $ I_R $ as its left and right children; \label{util_line}\\
		Send the counting query $ \qq_{I_L} $ to $ \Mc $ ;\\
		(By $ \Mc $) Respond to $ \Ac $ by $ \qq_{I_L}(S)$  ;\\
		Set $ n_{I_L} =  \qq_{I_L}(S)$ and $ n_{I_R} = n_{I_{\prdx}} - n_{I_L} $ ; \\
		Set $ \alphat_{I_{\prdx}} = \alpha_{I_{\prdx}} + n_{I_L}  $ and $ \betat_{I_{\prdx}} = \beta_{I_{\prdx}} + n_{I_R}  $ ;\\ 
		$ \Cc \leftarrow \Cc \cup \lrbrc{I_L,I_R} - \lrbrc{I_{\prdx}} $ ;\\
		$ i \leftarrow i+1 $ ;	
	}
\end{algorithm}

\section{Simulation studies} \label{S_Simul}
\subsection{Simulation scenario}
\subsubsection{Underlying distribution}
To assess the effectiveness of our proposed ADA scheme for PDF estimation, we assume that $ \Mc $ possesses $n$ samples drawn from a Gaussian mixture
\[ \text{GM}(x) = \frac{1}{M}\sum_{m=1}^{M}{\Nc(x;\mu_m,\sigma^2_m)} . \] 
The component parameters are generated independently and uniformly at random as follows:
\[ \mu_m \sim \Uc( -\mu_{\text{lim}} , \mu_{\text{lim}}  )  ,  \]
\[ \sigma_m \sim \Uc( \sigma_{\text{low}} , \sigma_{\text{high}}  ) .   \]
We assume that $x \in \mathcal{X} = I_0 = \left[-x_{\text{lim}}, x_{\text{lim}}\right]$. In the following, we set $M = 8$, $\mu_{\text{lim}} = 12.5$, $\sigma_{\text{low}} = 0.1$, $\sigma_{\text{high}} = 4$, and $x_{\text{lim}} = 25$. With these choices, the probability that each sample lies within $I_0$ is greater than $0.999$. Also, with this number of mixtures, we expect that in each realization, both narrow and wide peaks will be present with high probability. More precisely, if we denote the event that at least one component has $\sigma < 1$ as $A$, and the event that at least one component has $\sigma > 3$ as $B$, we have:
\begin{align}
	P(A \cap B) &= P(A) + P(B) - P(A \cup B) \nonumber \\
	& = 1- (3/3.9)^8 + 1 - (2.9/3.9)^8 -1 + (2/3.9)^8 \nonumber \\
	& \approxeq 0.788 \nonumber.
\end{align}
\subsubsection{Hyperparameters tuning}
As we noted before, the first step of the algorithm lets $ \Ac $ to promptly apply his intuitions about $ P $ in a hierarchical manner by deciding about $ (\rho,\eta) $ hyperparameters. As it's a standard routine in PTs literature, in this work we focus on $ \eta $ and assume that according to $ \Ac $'s prior beliefs, left and right child nodes in different levels have equal chances to inherit probability mass from their parent. Besides, in the literature, it is well-established that choosing $\eta_{ls} = \eta_{l} = cl^p$, with $p = 2$, generates random density functions that are almost surely continuous \cite{castillo2017polya}. Conversely, setting $p = -2$ recovers the Dirichlet process as a special case of Pólya Trees, which are known to produce fully discrete distributions. In general, the larger the value of $p$, the smoother the randomly generated densities. In this study, we assume that $\Ac$ takes this fact into account when evoking his prior belief about the smoothness of $P$ in different resolutions. Besides, gradual progress of $ \Ac $ in estimation lets him for more adjustment of $ \eta $ during the process, according his inspections. In this study, we assume that $ \Ac $ has only basic knowledge about $ P $. Specifically, $ \Ac $ knows that $ P $ is continuous and believes in a lower bound on the width of its peaks. In other words, he doesn't expect spiky behavior in $ P $ when the spikes are excessively narrow. We assume that $ \Ac $'s knowledge about this minimum width is aligned with $ \sigma_{\text{low}} $. We show that even invoking such partial information preserves the analysis from the curse of the ``garden of forking paths". At the same time, $ \Ac $ benefits from the power of adaptivity, which reduces the number of queries needed to achieve a desired level of accuracy, thereby enjoying the blessing of ``paradise of forking paths". To accommodate his belief about the shape of $ P $, we suggest that $ \Ac $ partially modifies the quadratic growth of $ \eta_{I} $ as follows:
\begin{equation} \label{eta_modif}
	\eta_{ls} = \eta_{l}(i) = 
	\begin{cases}
		0.4 \cdot (1.02)^i l^2  & \text{if } x \leq l_{\max} , \\
		0.4 \cdot (1.02)^i l^2 ( 1 + (n/2)/l_{\max}^2 ) & \text{if } x > l_{\max},
	\end{cases}
\end{equation}
where $ l_{\max} = \ceil{\log_2 (x_{\text{lim}} / \sigma_{\text{low}})} $ represents the finest resolution meaningful to him, and $ i $ denotes the round during which $ \prdx(l,s) $ is queried and subsequently split into $ (l,s) $ and its sister. 

We heuristically added a heavy penalty term to normally limit adaptive queries from exceeding this maximum depth until all candidates at this depth have been queried. Furthermore, the term $ (1.02)^i $ in \eqref{eta_modif} assigns a mild priority to candidate nodes that are added to $ \Cc $ earlier. This heuristic term differentiates between deeper nodes added to the list in earlier rounds and those added later. We emphasize that both of these modifications are intuitive, and their tuning is somehow heuristic. However, this is not a limitation; rather, it aligns perfectly with our goal of proposing a flexible framework for ADA, allowing $ \Ac $ to seamlessly incorporate his intuitive beliefs into analytic forms. Moreover, simulation results demonstrate that our suggestions are effective in density estimation task.

\subsubsection{Modification of connection points}
Considering the possibility of unequal lengths of adjacent intervals, we adjust the connection points between them to enhance the accuracy of density estimation. In detail, we define $ \ell_M \triangleq \max_{I \in T}{\ell_I}$, where $ \ell_I $ is the length of $ I $. Consider any two adjacent intervals, like $ T = [t_0, t_1)$ and  $ E = [e_0, e_1)$, where $ t_1 = e_0 $. Define $ o_{TE} \triangleq \min(\ell_T, \ell_ E)  $.
We adjust their junctions as follows:	
\begin{align}
	j_{T\rightarrow E} &= t_1 -  \lrpar{ \frac{\ell_T}{\ell_M} } \frac{o_{TE}}{2}  - \left( 1 - \frac{l_T}{\ell_M} \right) \frac{\ell_T}{2}, \\
	j_{E\rightarrow T} &= e_0 + \lrpar{\frac{\ell_E}{\ell_M}} \frac{o_{TE}}{2} + \left( 1 - \frac{\ell_E}{\ell_M} \right) \frac{\ell_E}{2}.
\end{align}
Intuitively, for longer intervals more shift is applied from their midpoint. This is done using a weighted average based on the comparison of the lengths of the intervals.

\subsection{Performance evaluation}
\subsubsection{Some examples}  
In the following, we refer to estimations with modified junctions as ``junc-ADA'', those with midpoint connections as ``mid-ADA'', and estimations thorough non-adaptive histograms as ``NADA''. To compare performance, first we visualize PDF estimations for three values of number of queries: $k = 25$, $80$, and $1000$, representing low, medium, and high numbers of queries, respectively, with $n =10^4$. Each estimation is derived by generating 30 random trees based on posteriorly updated PTs and node-wise averaging of random probability masses. 

\begin{figure}
	\centering
	\includegraphics[scale=0.4]{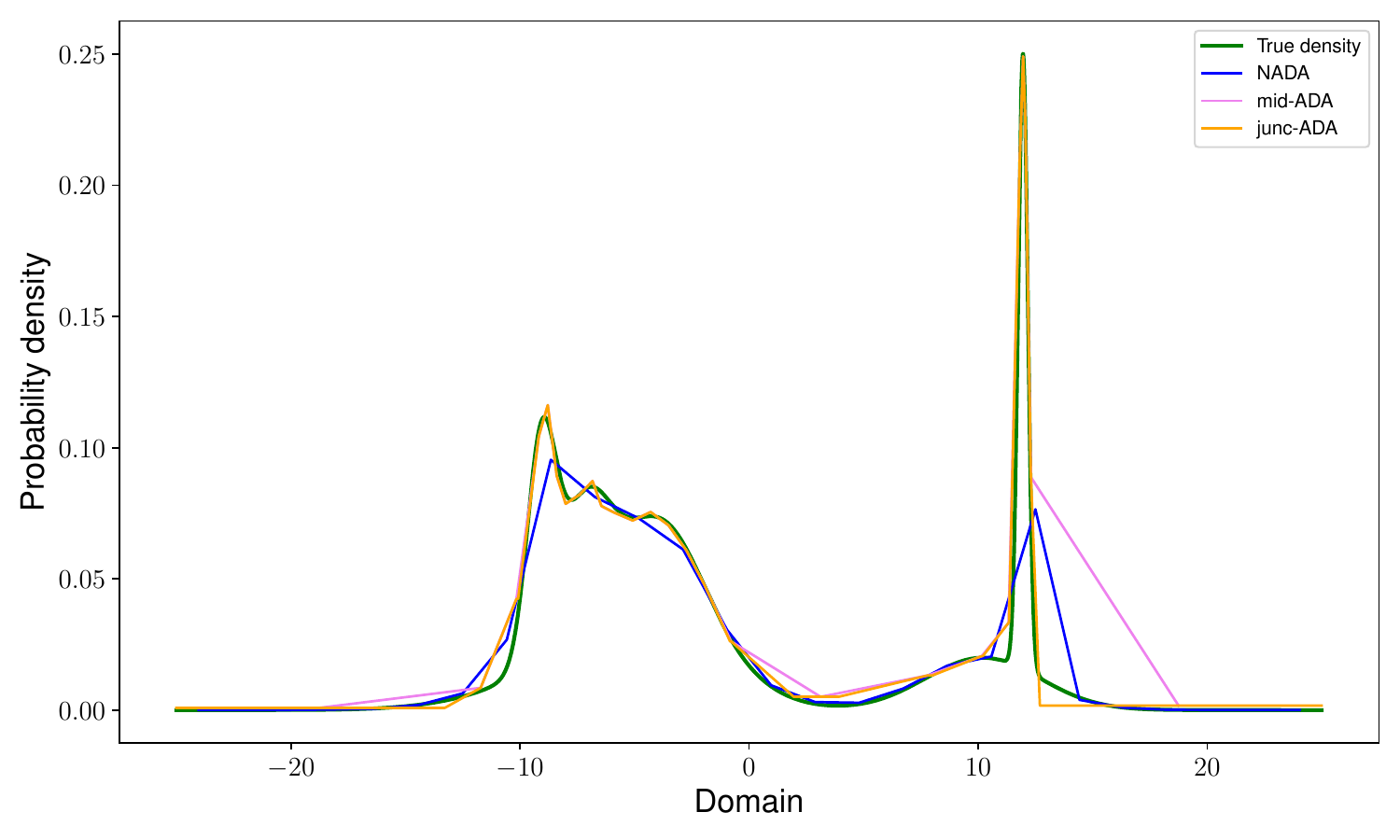}
	\caption{\label{fig:PDF_est_1} Comparison of the performance of NADA, mid-ADA, and junc-ADA in approximating the true density for $k=25$.}
\end{figure}

\begin{figure}
	\centering
	\includegraphics[scale=0.4]{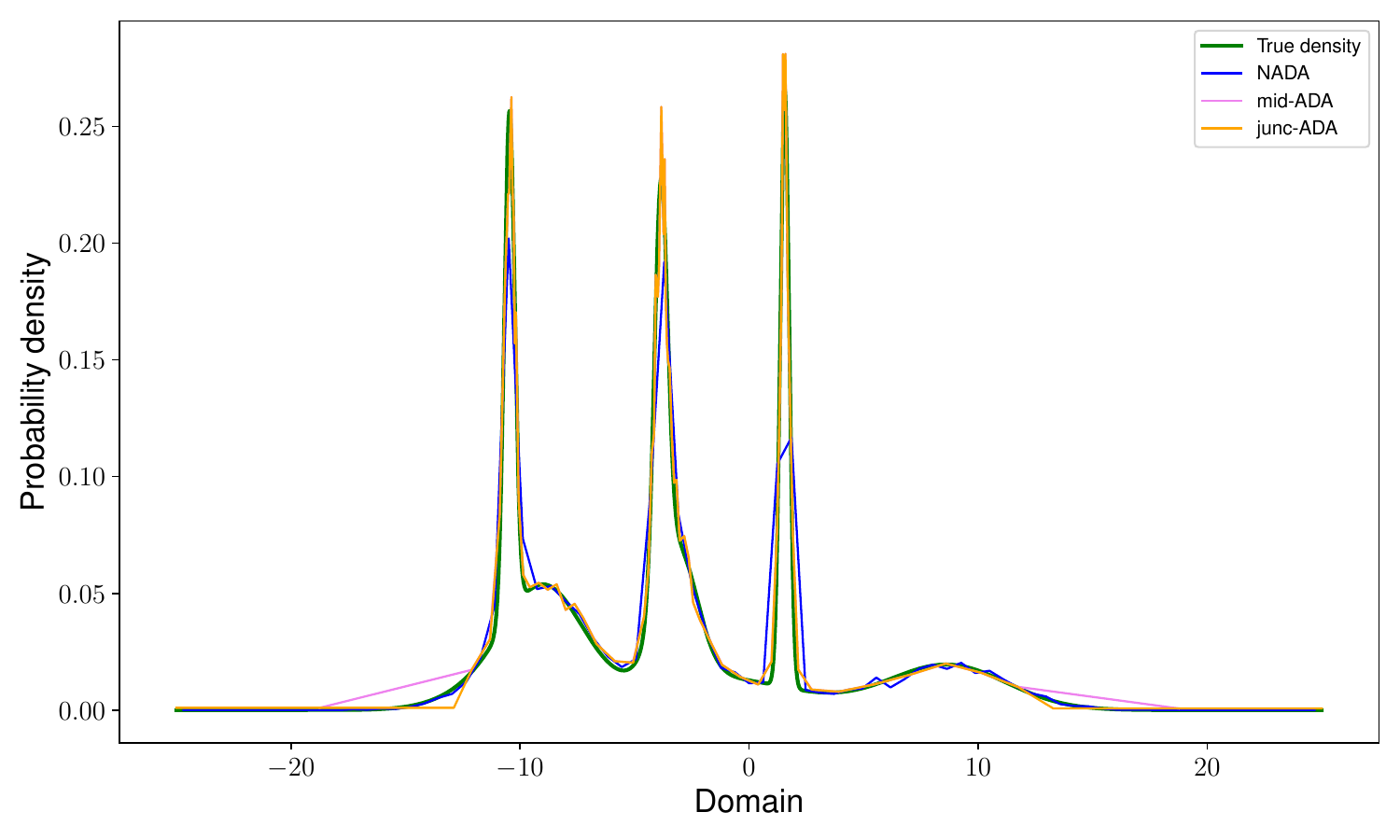}
	\caption{\label{fig:PDF_est_2} Comparison of the performance of NADA, mid-ADA, and junc-ADA in approximating the true density for $k=80$.}
\end{figure}

\begin{figure}
	\centering
	\includegraphics[scale=0.4]{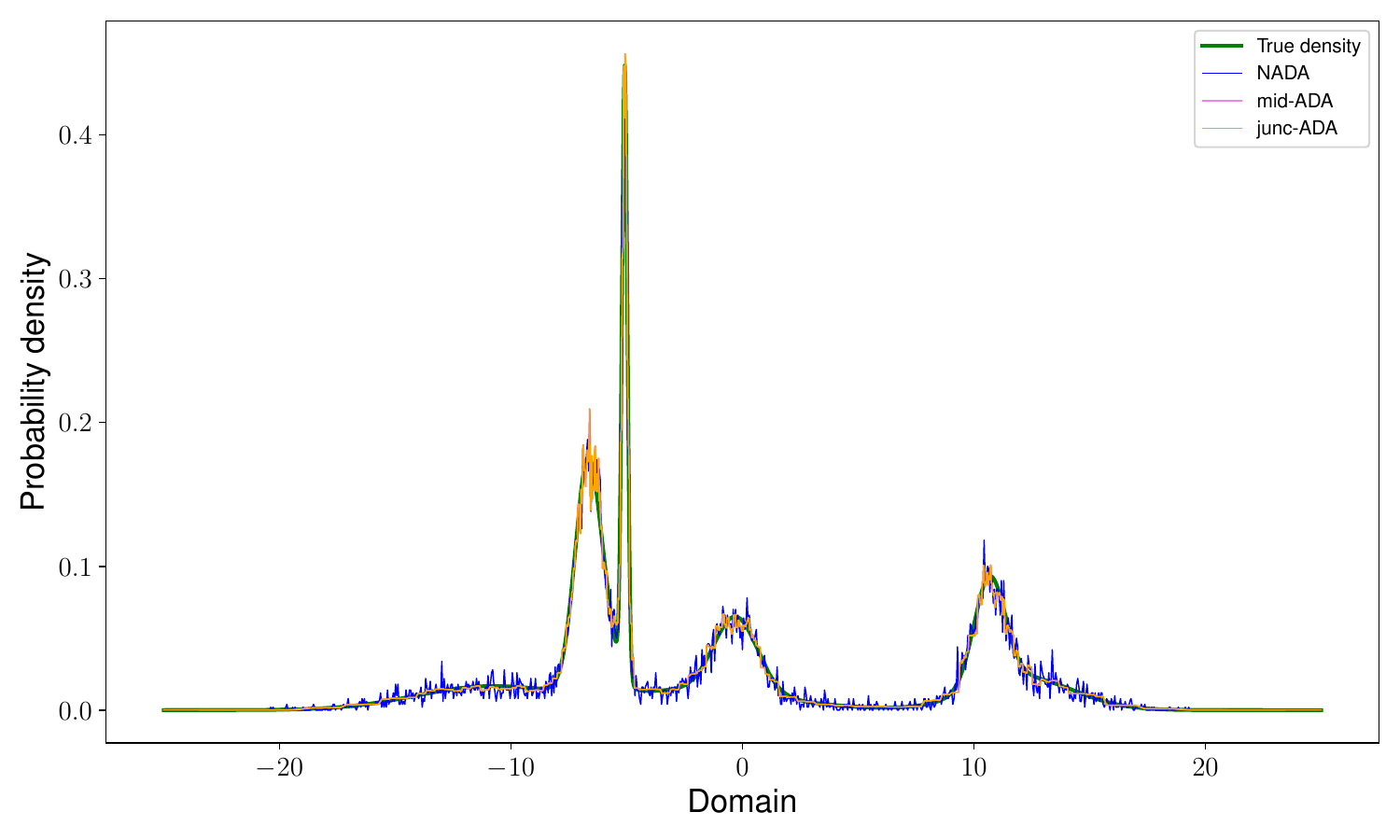}
	\caption{\label{fig:PDF_est_3} Comparison of the performance of NADA, mid-ADA, and junc-ADA in approximating the true density for $k=1000$.}
\end{figure}

In Fig. \ref{fig:PDF_est_1}, junc-ADA clearly outperforms NADA in both peak detection and capturing the overall shape. In Fig. \ref{fig:PDF_est_2}, junc-ADA highly succeeds to recover $P$, while $k$ remains insufficient for NADA to accomplish such a challenging task.  In Fig. \ref{fig:PDF_est_3}, although all the estimates follow $P$, the spikes in NADA tend
to surround those of mid-ADA and junc-ADA across most of the domain. Overall, we observe that junc-ADA provides a more accurate estimate than mid-ADA in regions where adjacent intervals have unequal lengths.

\subsubsection{Comparison of MSE and Total Variation distance} 
\begin{figure}
	\centering
	\includegraphics[scale=0.4]{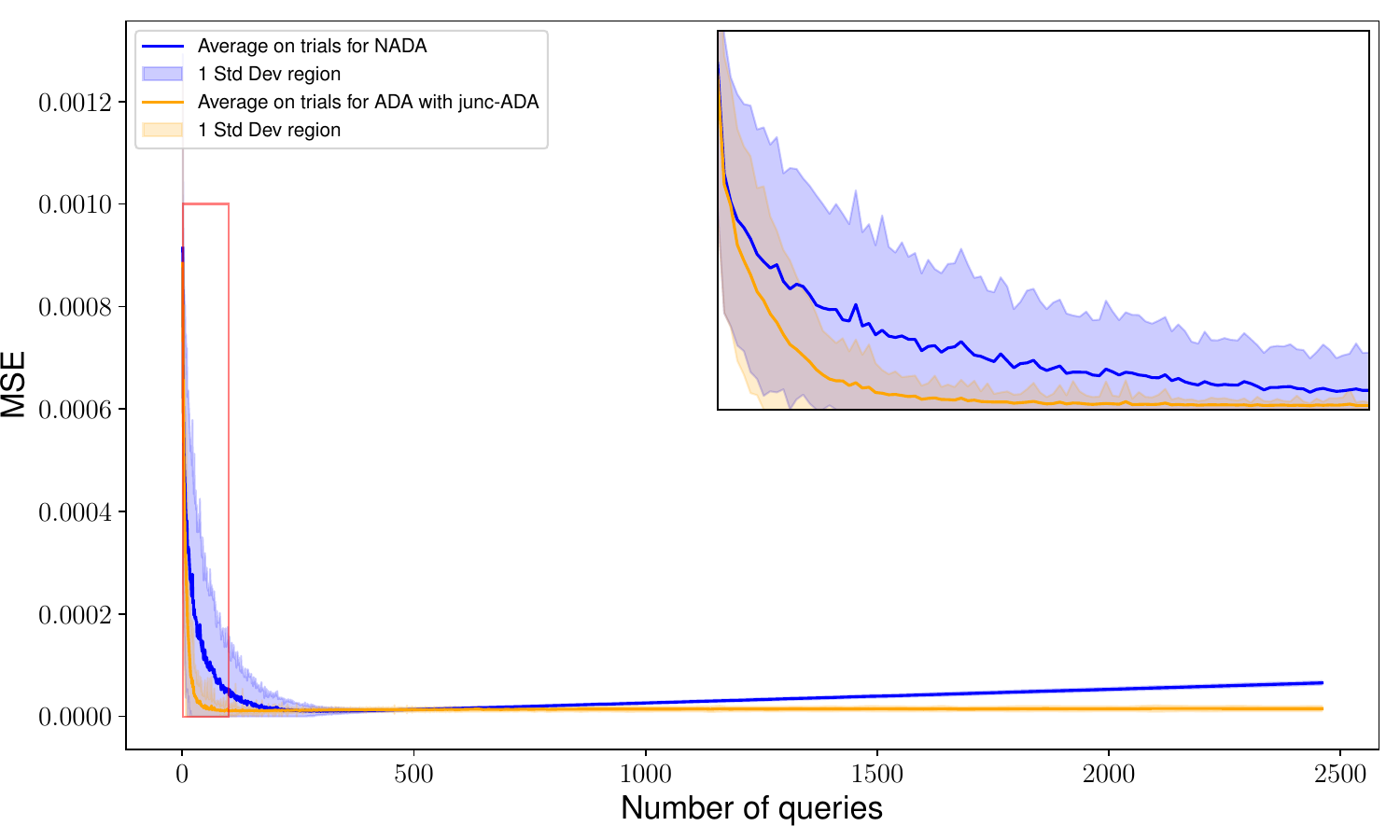}
	\caption{\label{fig:mean_MSE_compare_10e4} Comparing mean of MSE of trials.}
\end{figure}

\begin{figure}
	\centering
	\includegraphics[scale=0.4]{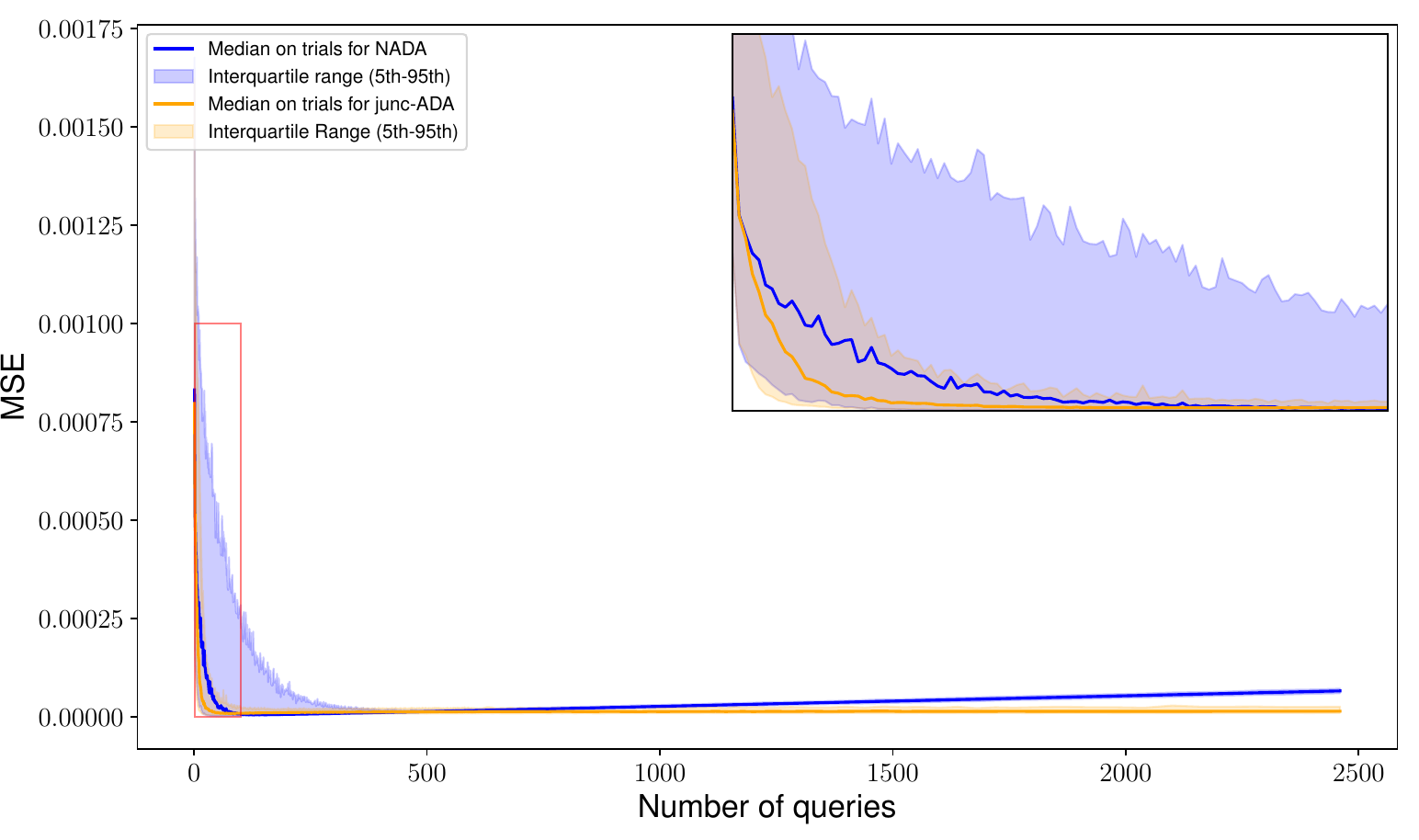}
	\caption{\label{fig:med_MSE_compare_10e4} Comparing median of MSE of trials.}
\end{figure}

\begin{figure}
	\centering
	\includegraphics[scale=0.4]{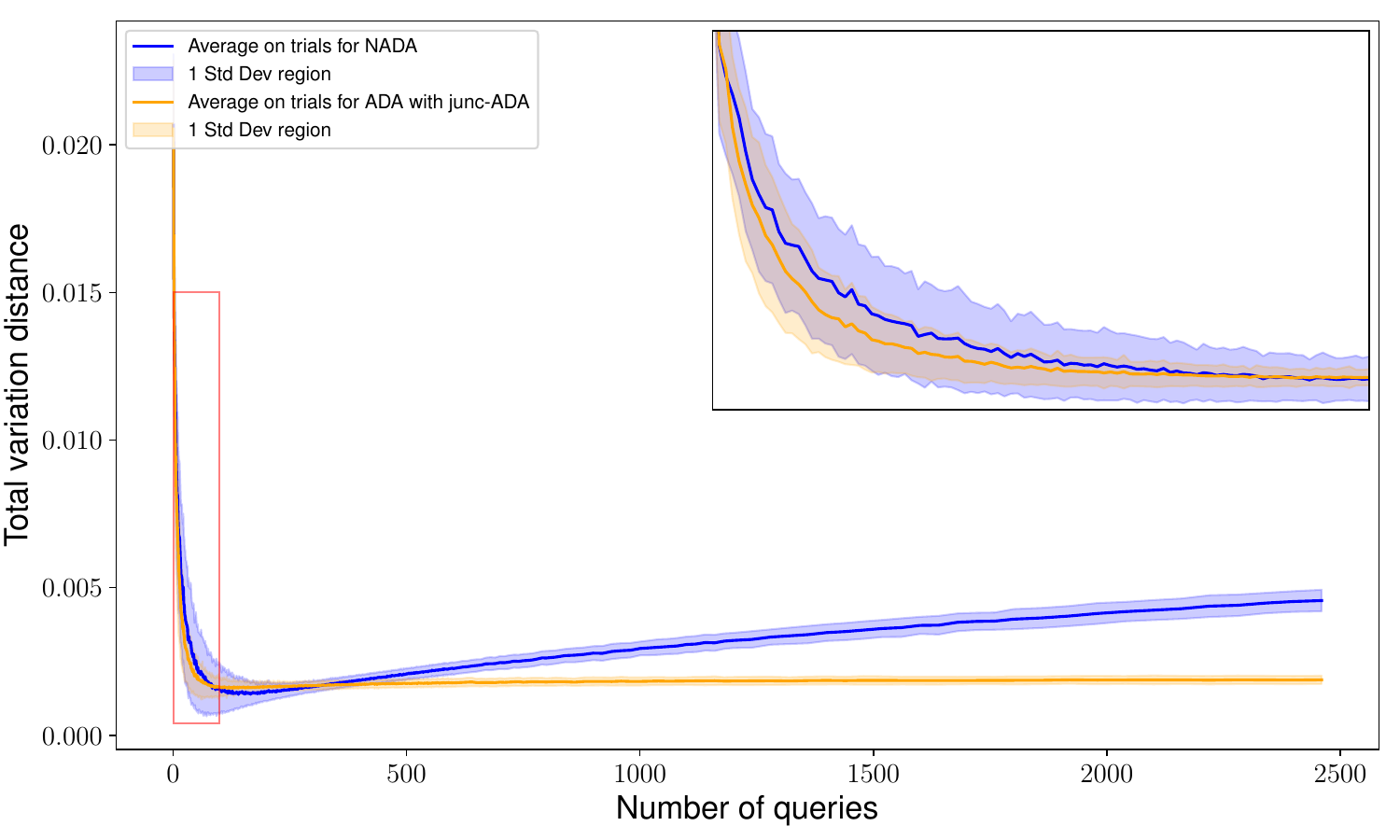}
	\caption{\label{fig:mean_TV_compare_10e4} Comparing mean of TV error of trials.}
\end{figure}

Here, we compare the performance of junc-ADA and mid-ADA for $ n =10^4 $ and $ k \in [1,2500) $, with  450 trials for each $ k $, and using the mean and median statistics of the trials. We also display the region around the mean corresponding to one standard deviation at each point. For the median plots, we highlight the interquartile range, showing the region between the 5th and 95th percentiles.  For better comparison, zoomed-in versions of the initial sections of each plot are provided at the top-right. Figure \ref{fig:mean_MSE_compare_10e4} demonstrates that the mean of MSE of trials for junc-ADA quickly reaches its best values after $k = 25$, whereas for NADA, this value remains above the one standard deviation region of junc-ADA for small and medium values of $k$. After $ k = 500 $, NADA begins to overfit, while junc-ADA largely maintains its performance. Fig. \ref{fig:med_MSE_compare_10e4} shows that, with high probability, the MSE performance of junc-ADA  in any trial is better than the median MSE of NADA for $ k $ about $ [25,100] $. 
These two figures show desirable performance of  junc-ADA in MSE, the metric for which the method is designed. By this metric, the method is encouraged to give priority to detecting the peaks. 
But what about the overall shape? Interestingly, Figs. \ref{fig:med_MSE_compare_10e4} and \ref{fig:mean_TV_compare_10e4} reveal that junc-ADA excels NADA in total variation (TV) distance for small values of $ k $. For medium values of $ k $, junc-ADA performs comparably to NADA, though with slightly weaker performance. Furthermore, although NADA overfits after $ k=300 $, junc-ADA approximately preserves its performance. 

\begin{figure}
	\centering
	\includegraphics[scale=0.4]{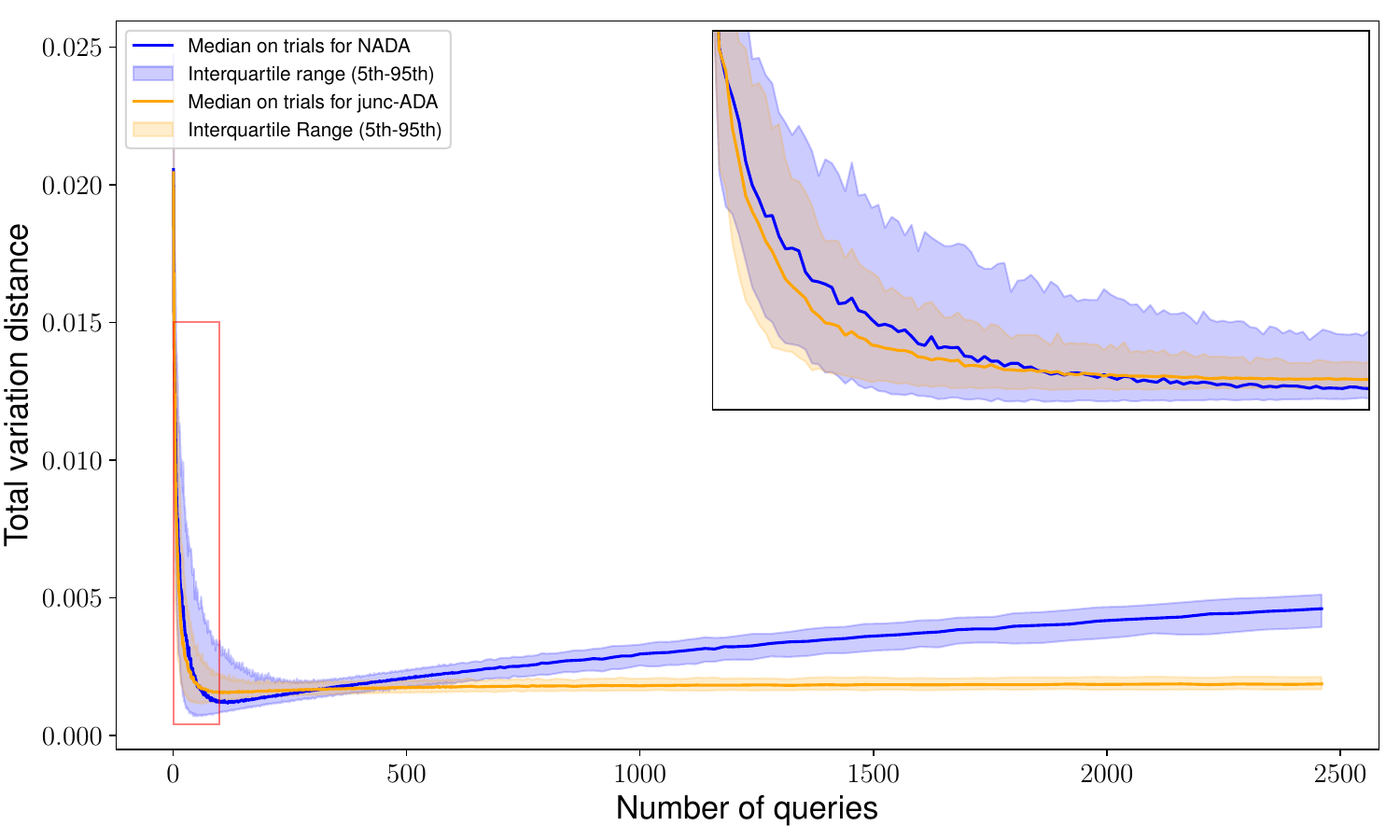}
	\caption{\label{fig:med_TV_compare_10e4} Comparing median of TV error of trials.}
\end{figure}



\section{Conclusions} 
\label{S_Concul}
Interacting with AI answering systems exemplifies a scenario where a human, leveraging subjective information, raises adaptive queries to a machine equipped with objective information and computational power. Asking queries in a divide-and-conquer manner represents a typical adaptive approach, aligning well with the constructive ADA model presented and analyzed in our paper.

Our scheme provides an intuitive framework for translating subjective beliefs into objective priors and adaptively utilizing previous findings to select informative queries hierarchically. This approach ensures not only strong statistical performance but also high interpretability, making our solution suitable for mental computations, either independently or with the aid of versatile computing assistants.

Based on the current study, we identify two promising areas for future research: first, extending the framework beyond counting queries; and second, refining the approximations used in the analysis of the second step of the algorithm to enhance accuracy.


%

\bibliographystyle{IEEEtran}
\bibliography{PhD}

\end{document}